\documentclass[final,5p,times,twocolumn]{elsarticle}

\usepackage{graphicx}

\usepackage{amssymb}

\journal{Computer Physics Communications}

\begin{document}

\begin{frontmatter}



\title{Efficient computation of Hamiltonian matrix elements between
 non-orthogonal Slater determinants
}


\author[label1,label2]{Yutaka Utsuno\corref{cor1}}
\ead{utsuno.yutaka@jaea.go.jp}
\author[label2]{Noritaka Shimizu}
\author[label2,label3,label4]{Takaharu Otsuka}
\author[label2]{Takashi Abe}
\address[label1]{Advanced Science Research Center, Japan Atomic Energy Agency,
Tokai, Ibaraki 319-1195, Japan}
\address[label2]{Center for Nuclear Study, University of Tokyo, Hongo Tokyo 113-0033, Japan}
\address[label3]{Department of Physics, University of Tokyo, Hongo,
 Tokyo 113-0033, Japan}
\address[label4]{National Superconducting Cyclotron Laboratory, Michigan
 State University, East Lansing, MI 48824, USA}
\cortext[cor1]{Corresponding Author}

\begin{abstract}

We present an efficient numerical method for computing Hamiltonian matrix elements  
between non-orthogonal Slater determinants, focusing on the most
time-consuming component of the calculation 
that involves a sparse array. 
In the usual case where many matrix elements should be calculated, 
this computation 
can be transformed into a multiplication of 
dense matrices. 
It is demonstrated that the present method 
based on the matrix-matrix multiplication attains $\sim$80\% of 
the theoretical peak performance measured on systems equipped with
modern microprocessors, 
a factor of 5-10 better than the normal method using indirectly
indexed arrays to treat a sparse array. 
The reason for such different performances is discussed 
from the viewpoint of memory access. 

\end{abstract}

\begin{keyword}
%
%
quantum many-body problem \sep Hamiltonian overlap \sep BLAS
\end{keyword}

\end{frontmatter}


\section{Introduction}
\label{}

One of the main issues in the quantum many-body  problem is 
solving a Schr\"{o}dinger equation to good accuracy 
in reasonable computational time. 
While mean-field methods such as the Hartree-Fock 
method are very successful in various systems, the inclusion of 
effects beyond the mean field, i.e., correlation, is highly desired 
for better description. 
For instance, the mean-field 
wave function does not necessarily have a good quantum number 
that is conserved in the exact solution such as the 
total angular momentum. 

A superposition of Slater determinants is the usual way to 
overcome the limitation of the mean-field method. 
Among various schemes to represent a correlated wave function, 
a representation by non-orthogonal Slater determinants 
(or quasiparticle vacuum states in general) 
is a method which is widely used in the nuclear many-body problems 
\cite{ring}. 
This method, often associated with the generator coordinate method (GCM) \cite{gcm}, 
has been successfully applied, for instance, 
to the description of collective motion 
and to the restoration of broken symmetry \cite{bender03}. 
Recently, global studies of the correlation energy and the 
energy spectra over the nuclear chart 
have been carried out with the use of 
the GCM, for instance in \cite{bender06, sabbey07, robledo11}. 
Furthermore, the use of non-orthogonal Slater determinants has recently opened 
a new possibility for representing a precise many-body wave 
function in an efficient way, 
as demonstrated by the Monte Carlo shell model (MCSM) \cite{mcsm}, 
variants of the VAMPIR method \cite{vampir}, 
and a hybrid method between MCSM and  VAMPIR \cite{puddu}. 
The MCSM method is now 
capable of precisely evaluating the eigenvalues 
even for a system beyond exact calculation 
by introducing a novel extrapolation method utilizing the variance 
of energy \cite{extrapolation}.  
There have been some studies using the superposition of 
non-orthogonal Slater determinants also in quantum chemistry 
\cite{koch, tomita, morgon, scuseria}. 

In the present paper, in order to extend the applicability 
of the expression of non-orthogonal Slater determinants, 
we present a numerical method 
for efficiently computing Hamiltonian matrix elements between them. 
Since we assume a general two-body force that 
has the rotational symmetry only, 
the present method will be applicable to various systems. 
This paper is organized as follows. Section \ref{sec:system} briefly
describes the many-body system and many-body wave function 
under consideration. 
Section~\ref{sec:numerical} presents 
some numerical methods for computing the most time-consuming 
part. In Sec.~\ref{sec:measurement}, the computational 
performances of the presented methods 
are compared, and the reason for 
their differences in performance is discussed. 
In Sec.~\ref{sec:summary}, we summarize this paper. 

\section{Many-body calculation with non-orthogonal Slater determinants}
\label{sec:system}

In this paper, we consider the many-body system described by the Hamiltonian 
consisting of a one-body operator $T$ and a two-body operator 
$V$, 
\begin{equation}
H=T + V = \sum_{l_1 l_2}^{N_{s}}t_{l_1l_2}c^{\dag}_{l_1}c_{l_2}+\frac{1}{4}
\sum_{l_1 l_2 l_3 l_4}^{N_{s}}\bar{v}_{l_1l_2,l_3l_4}
 c^{\dag}_{l_1}c^{\dag}_{l_2}c_{l_4}c_{l_3}, 
\label{eq:hamiltonian}
\end{equation}
where $c^{\dag}_l$ and $c_l$ are the creation and annihilation 
operators of the state labeled by $l$, respectively. 
The one-body matrix elements $t_{l_1 l_2}$ are 
given by $t_{l_1 l_2} = \langle l_1 | T | l_2 \rangle$, 
and the two-body matrix elements defined by 
$\bar{v}_{l_1l_2,l_3l_4} = \langle l_1 l_2 | V | l_3 l_4 \rangle - 
\langle l_1 l_2 | V | l_4 l_3 \rangle$ 
are antisymmetrized: 
$\bar{v}_{l_1l_2,l_3l_4} =- \bar{v}_{l_1l_2,l_4l_3}$. 
We consider a model space consisting of a finite number of 
single-particle orbits represented by $N_s$, and regard 
a set of the single-particle wave functions $\phi_l(x)=\langle x | c^{\dag}_l|- \rangle$ 
($l=1, 2, \ldots, N_s$)
as a {\it single-particle} basis set.  

We approximate the solution of Eq.~(\ref{eq:hamiltonian})
by a superposition of 
a finite number of non-orthogonal Slater determinants  
\begin{equation}
\left| \Psi \rangle\right. = 
\sum_q f(q) \left| \Phi(q) \rangle\right., 
\label{eq:wf}
\end{equation}
where $\left| \Phi(q) \rangle\right.$ and $f(q)$ 
denote a Slater determinant and its amplitude, respectively. 
Note that 
although the wave function $| \Psi \rangle$ is sometimes expressed 
by a continuous superposition over $q$ 
as is expressed by the GCM, 
the actual numerical calculation is usually performed 
by the discretization shown in Eq.~(\ref{eq:wf}). 
Each Slater determinant, regarded as a {\it many-body} basis state, 
is represented by a product of 
generalized creation operators 
\begin{equation}
\left| \Phi(q) \rangle\right. = \prod_{i=1}^{N_p} a^{\dag}_i(q) \left| -
 \rangle\right. ,
\label{eq:slater}
\end{equation}
where $N_p$ is the number of particles, and the creation operator 
$a^{\dag}_i(q)$ is given by 
\begin{equation}
\label{eq:occupied}
a^{\dag}_i(q) = \sum_{l}^{N_s}D(q)_{li}c^{\dag}_l . 
\end{equation}
Here the $N_s \times N_p$ matrix ($N_s\ge N_p$) $D(q)$ 
characterizes the many-body basis state 
$\left| \Phi(q)\rangle\right.$. 
In general, the basis states $\left| \Phi(q)\rangle\right.$ 
are non-orthogonal between one another: $\langle \Phi(q') | \Phi(q)
\rangle \ne 0$. 
Although an important issue in quantum many-body theory is how to choose 
good $\left| \Phi(q)\rangle\right.$, 
we do not mention it here because the aim of this paper 
is to present an efficient computational method which is 
valid for any calculation of the same type. 
Once a set of the many-body basis states is fixed, one needs to 
optimize a set of amplitudes $f(q)$. 
This optimization is usually carried out with the variational principle: 
\begin{equation}
\delta \frac{\langle \Psi \left| H \right| \Psi \rangle}
{\langle \Psi | \Psi \rangle} = 0, 
\end{equation}
which leads to the Hill-Wheeler equation \cite{gcm} for a discretized
coordinate $q$: 
\begin{equation}
\mathcal{H}f=E\mathcal{N}f, 
\label{eq:hw}
\end{equation}
where $\mathcal{H}$ and $\mathcal{N}$ are matrices
whose elements are given by 
\begin{eqnarray}
\mathcal{H}(q',q) &=& \langle \Phi(q') \left| H \right| \Phi(q) \rangle\\
\mathcal{N}(q',q) &=& \langle \Phi(q') | \Phi(q) \rangle .
\end{eqnarray}
$f$ is a vector whose component is $f(q)$, 
and $E$ is the eigenvalue. 
Following the terminology of the GCM, we hereafter call 
the {\it many-body} matrix elements of 
$\mathcal{H}$ and $\mathcal{N}$ 
the Hamiltonian overlap and the norm overlap, 
respectively, 
to avoid confusing them with the two-body matrix element 
$\bar{v}_{l_1l_2,l_3l_4}$ of a single-particle basis. 
Both the overlaps are represented by $D(q)$ and $D(q')$. 
The norm overlap is written as 
\begin{equation}
\label{eq:norm_kernel}
\mathcal{N}(q',q) = \det\left( D(q')^{\dag}D(q)\right) , 
\end{equation}
and the Hamiltonian overlap is 
\begin{equation}
\label{eq:hamiltonian_kernel}
\begin{array}{l}
\displaystyle{ \mathcal{H}(q',q)
=  \mathcal{N}(q',q) \left( 
		      \sum_{l_1l_2}^{N_s}t_{l_1l_2}\rho_{l_2l_1}\right. }\\ 
\displaystyle{
\left. +\frac{1}{2}
\sum_{l_1l_2l_3l_4}^{N_s}\rho_{l_3l_1}\bar{v}_{l_1l_2,l_3l_4}\rho_{l_4l_2}
\right) }
\end{array}
\end{equation}
using the density matrix $\rho$ whose matrix element is defined by 
\begin{equation}
\label{eq:density}
\rho_{ll'} =
\frac{\langle \Phi(q') | c^{\dag}_{l'} c_{l}| \Phi(q)
\rangle}
{\langle \Phi(q') | \Phi(q) \rangle}. 
\end{equation}
Using $D(q)$ and $D(q')$, the density matrix becomes 
\begin{equation}
\label{eq:rho}
\rho = D(q)\left( D(q')^{\dag}D(q)\right)^{-1}D(q')^{\dag} .
\end{equation}
The derivation of Eqs.~(\ref{eq:norm_kernel}),
(\ref{eq:hamiltonian_kernel}) and (\ref{eq:rho}) is given in 
\ref{app:kernel}. 

Among various applications of the above expression 
is the restoration of broken symmetries. 
Since a general Slater determinant of Eq.~(\ref{eq:slater}) 
does not necessarily possess the symmetries that the original Hamiltonian 
has, it is desirable to restore the broken symmetries by projecting the wave function 
onto good quantum numbers. 
The total angular momentum, for instance, is 
restored from $| \Phi \rangle$ 
by performing a three-dimensional integration over the Euler angles 
\cite{ring}.  To carry out a numerical integration, 
the number of mesh points for the Euler angles 
is required to be as many as the order of $10^4$, 
as are the numbers of $\mathcal{H}(q',q)$ and $\mathcal{N}(q',q)$ 
to be calculated \cite{ring}. 

As thus exemplified,  
innumerable Slater determinants are often involved to obtain a good 
many-body wave function $| \Psi \rangle$. 
Hence, 
fast computation of the Hamiltonian and norm overlaps will  
accelerate the whole calculation. 
The most time-consuming in the above procedure
is the computation of the two-body part of the Hamiltonian overlap
\begin{equation}
\begin{array}{rcl}
\langle V \rangle & \equiv &
\displaystyle{
\sum_{l_1l_2l_3l_4}^{N_s}\rho_{l_3l_1}\,\bar{v}_{l_1l_2,l_3l_4}\,\rho_{l_4l_2}} \\
& = & 
\displaystyle{
\sum_{l_1 l_3} ^{N_s}\rho_{l_3l_1}\,\Gamma_{l_1 l_3}
}, 
\end{array}
\label{eq:vcal}
\end{equation}
with
\begin{equation}
\Gamma_{k k'} = \sum_{l l'}^{N_s}
 \bar{v}_{kl',k'l}\,\rho_{l l'}, 
\end{equation}
because such computation requires a fourfold summation over the single-particle states. 
In the following sections, we concentrate on an 
efficient computational method for Eq.~(\ref{eq:vcal}) 
on systems equipped with modern microprocessors. 
We assume that 
the operation of Eq.~(\ref{eq:vcal}) is repeated 
a great number of times for different density matrices $\rho$ 
under the condition of fixed two-body matrix elements $\bar{v}_{l_1l_2,l_3l_4}$.  

\section{Numerical methods for computing the Hamiltonian overlap}
\label{sec:numerical}

A straightforward operation of Eq.~(\ref{eq:vcal}) is in general 
a waste of computational time because $\bar{v}_{l_1l_2,l_3l_4}$ is very sparse. 
This sparseness is due to the symmetry of the Hamiltonian. For instance, 
the conservation of the $z$ component of the angular momentum 
leads to $\bar{v}_{l_1l_2,l_3l_4}=0$ 
unless $j_z(l_1)+j_z(l_2)=j_z(l_3)+j_z(l_4)$ is satisfied. 
Depending on the system considered, 
some other symmetries such as 
parity, orbital angular momentum,  
and isospin quantum numbers are also conserved, which imposes further 
constraints on the non-zero matrix elements. 
Hence, every effort must be made to avoid taking 
those vanishing matrix elements for efficient computing. 
Below we show three numerical algorithms for this purpose. 
The first method is completely different from the other two, and 
the last method is more advanced than the second method. 

\subsection*{Indirect-index method}

As shown in the last paragraph, 
the operation associated with zero for calculating $\langle V \rangle$ 
is mainly caused not by 
the density matrix but by the fixed two-body matrix elements
$\bar{v}_{l_1l_2,l_3l_4}$. 
Thus, it is useful to classify in advance  
the indices $(l_1, l_2, l_3, l_4)$ of $\bar{v}_{l_1l_2,l_3l_4}$ 
according to whether 
they lead to non-vanishing $\bar{v}_{l_1l_2,l_3l_4}$, and to label the 
set of indices $(l_1, l_2, l_3, l_4)$ 
satisfying this condition 
with a so-called indirect index $k$ 
as $(l_1(k), l_2(k), l_3(k), l_4(k))$. 
Equation~(\ref{eq:vcal}) is then represented as 
\begin{equation}
\langle V \rangle =
\sum_{k}^{N_{\rm{nonzero}}} \rho_{l_3(k)l_1(k)}
\,\bar{v}_{\rm{nonzero}}(k) \,
\rho_{l_4(k)l_2(k)}, 
\label{eq:ind}
\end{equation}
where $N_{\rm{nonzero}}$ is the number of non-vanishing 
$\bar{v}_{l_1l_2,l_3l_4}$, and 
$\bar{v}_{\rm{nonzero}}(k) \equiv \bar{v}_{l_1(k)l_2(k),l_3(k)l_4(k)}
\ne 0$. 
When $\bar{v}_{l_1l_2,l_3l_4}$ is sparse, $N_{\rm{nonzero}}$ 
is much smaller than $N_s^4$. 
In this paper, we refer to the numerical algorithm based on 
Eq.~(\ref{eq:ind}) as the {\it indirect-index method}.

\subsection*{Matrix-vector method}

Although the introduction of the indirect index can always be 
applied to the computation of sparse arrays, here 
we present an alternative numerical approach which directly 
utilizes the symmetry. We now assume that the two-body 
force $V$
has only the rotational invariance for simplicity. 
Other possible symmetries can be treated in a similar way. 

First, $N_s\times N_s$ density-matrix elements $\rho_{ll'}$ are 
grouped according to $\Delta m \equiv j_z(l')-j_z(l)$, and 
the set of ($l, l'$) having a common $\Delta m$ 
is indexed by $k=1, 2, \ldots, N_{\Delta m}$ 
as $\tilde{\rho}(\Delta m)_k$. 
In a similar way, the two-body matrix elements $\bar{v}_{l_1l_2,l_3l_4}$ 
are categorized according to $\Delta m_{13} \equiv j_z(l_1)-j_z(l_3)$ 
and $\Delta m_{24} \equiv j_z(l_2)-j_z(l_4)$ as 
$\tilde{v}(\Delta m_{13}, \Delta m_{24})_{k'k}$, 
where $k'$ and $k$ are, respectively, indices to $(l_1, l_3)$ and ($l_2, l_4$) 
having $\Delta m_{13}$ and $\Delta m_{24}$. 
Equation~(\ref{eq:vcal}) then leads to 
\begin{equation}
\begin{array}{rcl}
\langle V \rangle  & = & 
\displaystyle{\sum_{\Delta m_{13} \Delta m_{24}} \sum_{k'k}
\tilde{\rho}(\Delta m_{13})_{k'} \tilde{v}(\Delta m_{13}, \Delta
m_{24})_{k'k}} \\
& & \times \displaystyle{
 \tilde{\rho}(\Delta m_{24})_{k}} \\
& = & \displaystyle{\sum_{\Delta m} \sum_{k'k}
\tilde{\rho}(-\Delta m)_{k'} \tilde{v}(-\Delta m, \Delta m)_{k'k}
 \tilde{\rho}(\Delta m)_{k}}, 
\end{array}
\label{eq:vmat}
\end{equation}
where the last equation of Eq.~(\ref{eq:vmat}) is derived from 
the necessary condition for $\bar{v}_{l_1l_2,l_3l_4}$ being non-zero: 
 $j_z(l_1)+j_z(l_2)=j_z(l_3)+j_z(l_4)$, i.e., 
$\Delta m_{13}= j_z(l_1)-j_z(l_3)=-(j_z(l_2)-j_z(l_4))=-\Delta m_{24}
\equiv - \Delta m$. 

\begin{figure}[t]
\begin{center}
\includegraphics[width=8.0cm,clip]{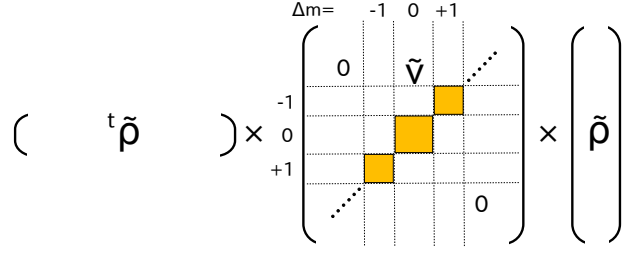}
\caption{Schematic illustration of the operation of
 Eq.~(\ref{eq:vmat}). 
}
\label{fig:matrix}
\end{center}
\end{figure}

Since the density matrix $\tilde{\rho}(\Delta m)$ 
and the two-body matrix $\tilde{v}(-\Delta m, \Delta m)$ 
for a given $\Delta m$ are
a one-dimensional
array and a two-dimensional array, respectively, 
they can be identified with 
a vector of size $N_{\Delta m}$ and a matrix of size 
$N_{\Delta m} \times N_{\Delta m}$, respectively,  
by using $N_{\Delta m} = N_{-\Delta m}$. 
Thus, 
Eq.~(\ref{eq:vmat}) is regarded as a 
$^t$(vector)$\times$(matrix)$\times$(vector) operation. 
This is schematically illustrated in Fig.~\ref{fig:matrix}. 
It is clearly seen that the sparse array $\bar{v}_{l_1l_2,l_3l_4}$ is 
transformed into a block-antidiagonal matrix $\tilde{v}$ whose 
blocks are dense submatrices. 
In this paper, we refer to the numerical algorithm based on
Eq.~(\ref{eq:vmat}) as the {\it matrix-vector method}.

\subsection*{Matrix-matrix method}

In the matrix-vector method, most of the computational time is devoted 
to the (matrix)$\times$(vector) operation 
$\tilde{\Gamma} \equiv \tilde{v}\tilde{\rho}$, 
where the index of $\Delta m$ is omitted for simplicity. 
As previously mentioned, this operation is usually repeated a number of times 
for different $\tilde{\rho}$'s: $\tilde{v}\tilde{\rho}^{(1)}, 
\tilde{v}\tilde{\rho}^{(2)}, \ldots$
By binding vectors $\tilde{\rho}^{(1)}, 
\tilde{\rho}^{(2)}, \ldots, \tilde{\rho}^{(N_{\rm vec})}$ into a matrix 
$\theta \equiv (\tilde{\rho}^{(1)},
\tilde{\rho}^{(2)}, \ldots, \tilde{\rho}^{(N_{\rm vec})})$, 
repeated (matrix)$\times$(vector) operations are 
performed by a (matrix)$\times$(matrix) operation at one time: 
\begin{equation}
(\tilde{\Gamma}^{(1)}, \tilde{\Gamma}^{(2)}, \ldots,
 \tilde{\Gamma}^{(N_{\rm vec})}) =
(\tilde{v}\tilde{\rho}^{(1)}, \tilde{v}\tilde{\rho}^{(2)}, \ldots, 
\tilde{v}\tilde{\rho}^{(N_{\rm vec})}) = 
\tilde{v}\theta, 
\label{eq:vtheta}
\end{equation}
where the number of columns $N_{\rm vec}$ can be chosen arbitrarily.   
The $\langle V \rangle$ for the $i$-th 
density matrix $\tilde{\rho}^{(i)}$ is then given by
$^t\tilde{\rho}^{(i)}\tilde{v}\tilde{\rho}^{(i)} = 
{^t\tilde{\rho}^{(i)}}\tilde{\Gamma}^{(i)} = 
{^t\tilde{\rho}^{(i)}}(\tilde{v}\theta)^{(i)}$, where
$(\tilde{v}\theta)^{(i)}$ stands for the $i$-th column of the 
matrix $\tilde{v}\theta$. 
We call this method, i.e, the way through the (matrix)$\times$(matrix) 
operation of Eq.~(\ref{eq:vtheta}), the {\it matrix-matrix method}.  
It seems as if there is no substantial difference between the matrix-matrix method 
and the matrix-vector method: equation (\ref{eq:vtheta})  
keeps not only mathematical identity 
but also the number of elementary operations. 
However, as seen in the next section, 
those two methods result in quite different computational performances 
on actual computer systems. 

\subsection*{Case of the quasiparticle vacuum state}

Although this paper concentrates on the Hamiltonian overlap 
between Slater determinants, it is useful to mention applicability 
to the Hamiltonian overlap between quasiparticle vacuum states. 
The quasiparticle vacuum state is a generalized single-particle state,
and is widely used in nuclear physics 
to include the pairing correlation. 
Similar to Eq.~(\ref{eq:hamiltonian_kernel}), the Hamiltonian 
overlap for the quasiparticle vacuum state is written \cite{ring} as 
\begin{equation}
\label{eq:hamiltonian_hernel_hfb}
\begin{array}{l}
\langle \Phi(q') | H | \Phi(q) \rangle  =  
\langle \Phi(q') | \Phi(q) \rangle  \\ 
\displaystyle{ \times
\left( \sum_{l_1l_2}^{N_s}t_{l_1l_2}\rho^{10}_{l_2l_1}   + 
\frac{1}{2}
\sum_{l_1l_2l_3l_4}^{N_s}\rho^{10}_{l_3l_1}
\bar{v}_{l_1l_2,l_3l_4}\rho^{10}_{l_4l_2} \right.
} 
\\
\displaystyle{ \left.
+ \frac{1}{4} 
\sum_{l_1l_2l_3l_4}^{N_s} \kappa^{01*}_{l_1 l_2}
\bar{v}_{l_1l_2,l_3l_4} \kappa^{10}_{l_3 l_4}
\right) 
} , 
 \end{array}
\end{equation}
where the density matrix $\rho^{10}$ 
and the pairing tensors $\kappa^{10}$ and $\kappa^{01*}$ are 
defined by 
\begin{eqnarray}
\rho^{10}_{ll'} & = & 
\frac{\langle \Phi(q') | c^{\dag}_{l'} c_{l}| \Phi(q)
\rangle}
{\langle \Phi(q') | \Phi(q) \rangle} \\
\kappa^{10}_{ll'} & = & 
\frac{\langle \Phi(q') | c_{l'} c_{l}| \Phi(q)
\rangle}
{\langle \Phi(q') | \Phi(q) \rangle} \\
\kappa^{01*}_{ll'} & = & 
\frac{\langle \Phi(q') | c^{\dag}_{l}c^{\dag}_{l'} | \Phi(q)
\rangle}
{\langle \Phi(q') | \Phi(q) \rangle} .
\end{eqnarray}
The difference from the Slater determinant is the addition of the last
term in Eq.~(\ref{eq:hamiltonian_hernel_hfb}). 
Its computation with the matrix-vector or matrix-matrix method is 
rather similar. The two-body matrix elements $\bar{v}_{l_1 l_2, l_3 l_4}$ 
are categorized according to $M_{12} \equiv j_z(l_1)+j_z(l_2)$ 
and $M_{34} \equiv j_z(l_3)+j_z(l_4)$ 
as $\tilde{w}(M_{12}, M_{34})_{k'k}$, where 
$k'$ and $k$ are indices to $(l_1, l_2)$ having $M_{12}$ and ($l_3, l_4$)
having $M_{34}$,
respectively. Since the necessary condition for
$\bar{v}_{l_1l_2,l_3l_4}$ being non-zero is 
$M_{12}=M_{34} $ $(\equiv M)$, the two-body matrix elements are 
block diagonalized as $\tilde{w}(M, M)_{k'k}$ 
each of which is a dense matrix. 
The pairing tensor can be regarded as a vector $\tilde{\kappa}$ in this representation. 
Thus, when the matrix-vector or matrix-matrix method is applied to 
the quasiparticle vacuum state, one needs to prepare two kinds of 
matrix representations of $\bar{v}_{l_1 l_2, l_3 l_4}$, $\tilde{v}$ and
$\tilde{w}$, the former and the latter of which act on the vector representation 
of the density matrix, $\tilde{\rho}$, and 
the vector representation of the pairing tensor,
$\tilde{\kappa}$, respectively. 


\section{Measurement of performance}
\label{sec:measurement}

In this section, computational performance is compared 
among the three methods presented in the last section 
by adopting a realistic many-body system and measuring 
the elapsed time to compute 
Eq.~(\ref{eq:vcal}) repeatedly. 

\subsection{Benchmark system}

Here we consider a nuclear many-body problem where 
protons and neutrons interact in a fixed model 
space. 
We adopt a set of the single-particle orbits 
consisting of five harmonic-oscillator major shells from 
harmonic-oscillator's 
quantum number $N_{osc}=0$ to 4:  
$0s_{1/2}$,  $0p_{3/2}$, $0p_{1/2}$,  $0d_{5/2}$, $0d_{3/2}$, 
$1s_{1/2}$, $0f_{7/2}$, $0f_{5/2}$, $1p_{3/2}$, $1p_{1/2}$,  
$0g_{9/2}$, $0g_{7/2}$,  $1d_{5/2}$, $1d_{3/2}$,  and $2s_{1/2}$. 
Thus, the number of the proton (neutron) single-particle states $N_s$
is 70.  
Here, the proton and neutron numbers are set to be two and two, 
respectively,  
but the number of particles is irrelevant to the computational 
time of Eq.~(\ref{eq:vcal}). 

The two-body part of the adopted Hamiltonian is an arbitrary one that has rotational, 
parity and time-reversal symmetries. 
Due to the rotational and time-reversal symmetries, all the matrix 
elements $\bar{v}_{l_1 l_2, l_3 l_4}$ can be real numbers \cite{bm}. 
Since we do not assume other 
symmetries such as an isospin, we calculate the proton-proton 
interaction part of Eq.~(\ref{eq:vcal}), the neutron-neutron part, 
and the proton-neutron part independently. 
For this system, the largest submatrix used in the matrix-vector 
(or matrix-matrix) method is of the size $390\times 390$, 
classified according to the $z$ component of the angular momentum 
and the parity. 

The wave function taken is a single Slater determinant 
with total angular-momentum and parity projection. 
Each single-particle state of the Slater determinant 
is assumed to be a pure proton or neutron state. 
The number of mesh points for the three Euler angles and 
that for the parity projector are $25^3$ and 2, respectively, 
leading to $25^3\times 2=31,250$ times the computations of
Eq.~(\ref{eq:vcal}) . 
Since a rotation of a wave function 
involves imaginary numbers \cite{ring}, the density matrix has to be complex. 

It would be useful to compare the number of elementary 
floating-point operations 
(addition and multiplication) among the three methods. 
Taking into account that 
the loop length of Eq.~(\ref{eq:vcal}) can be 
halved by using $\bar{v}_{l_1 l_2, l_3 l_4}
=\bar{v}_{l_2 l_1, l_4 l_3}$, the number of elementary 
floating-point operations becomes 
20,992,518 for the indirect-index method and 
10,365,224 for the matrix-vector method
and the matrix-matrix method. 
The former 
is almost the double of the latter as explained as follows. 
In the matrix-vector method, 
$(\tilde{v}\tilde{\rho})_{k'}
= \tilde{v}_{k'1}\tilde{\rho}_1 + \tilde{v}_{k'2}\tilde{\rho}_2 +\ldots$ 
is factored out 
of $\sum_{k'k} \tilde{\rho}_{k'} \tilde{v}_{k'k}\tilde{\rho}_k$ 
in the way $\sum_{k'} \tilde{\rho}_{k'}
(\tilde{v}_{k'1}\tilde{\rho}_1 + \tilde{v}_{k'2}\tilde{\rho}_2
+\ldots)$. 
This expression saves the number of multiplications,  
and more importantly, the reduced operations are  
the multiplication of {\it complex numbers} which costs 
as many as six floating-point operations. 

\subsection{Computational environment}

The computation is carried out as a single-threaded process 
on two different systems 
based on up-to-date scalar processors: one 
system is based on the Xeon X5570 processor with 
clock speed 2.93 GHz and the other is based on the SPARC64 VII 
processor with clock speed 2.5 GHz. 
Their theoretical peak 
performances per CPU core are 11.72 GFLOPS and 10 GFLOPS, 
respectively. 
Our code written in Fortran 90/95/2003 is compiled 
by Intel Fortran Compiler Version 11.1 for the Xeon system 
and by Fujitsu Fortran Compiler Driver Version 8.2 
for the SPARC64 system. 
The two-body matrix elements $\bar{v}_{l_1 l_2, l_3 l_4}$ and 
the density matrix elements $\rho_{l l'}$ are 
of double-precision. 
Matrix and/or vector calculations are coded to call 
the BLAS interface
(BLAS \cite{blas} is the de facto standard for the programming interface 
of basic linear algebra operations). 
We use optimized BLAS implementations: 
Intel Math Kernel 
Library (MKL) for the Xeon system and Fujitsu Scientific Subroutine
Library II  (SSL II)
for the SPARC64 system. 
The computational performance 
for executing Eq.~(\ref{eq:vcal}) 
is measured with the wall-clock time 
at a microsecond-level resolution, which is good enough 
for the present purpose. 

\subsection{Results and analyses}
\label{sec:result}

\begin{figure}[t]
\begin{center}
\includegraphics[width=7.5cm,clip]{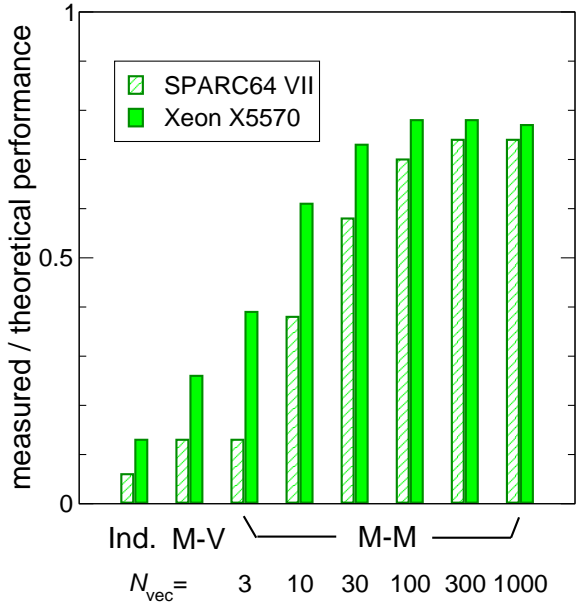}
\caption{
Comparison of the computational performance 
among the indirect-index method 
(Ind.), matrix-vector method (M-V) and  matrix-matrix method (M-M) 
with different $N_{\rm vec}$ measured on the 
SPARC64 VII and Xeon X5570 systems. 
The values are normalized by their theoretical peak performance. 
See the text for more details. 
}
\label{fig:performance}
\end{center}
\end{figure}

The performance of a computation is characterized by the inversion of 
the wall-clock time $t$. It is comprehensive to express the performance 
in FLOPS which is 
$t^{-1}$ in 1/second multiplied by the total number of elementary 
floating-point operations executed. 
But since the number of operations 
is different among the methods as shown previously,  
FLOPS is not a good measure for comparing their relative performances.  
Hence, to make direct comparison possible, 
the performance is now defined 
by $t^{-1}$  multiplied by a fixed factor of the number of 
elementary floating-point operations of the matrix-vector 
(or the matrix-matrix) method, only when 
it serves as the actual FLOPS. 

Figure~\ref{fig:performance} compares the measured performances  
normalized by the theoretical peak performances of the adopted systems. 
The indirect-index method gives the lowest performance 
for both systems. 
The performance of the matrix-vector method is about twice as high as 
that of the indirect-index method. This is almost equivalent to the ratio 
of the numbers of floating-point operations, 
but is still far from the theoretical peak performance. 
When $\tilde{\rho}$ vectors are bound into a matrix 
in the matrix-matrix method, the performance starts to increase. 
The performance improves sharply even at a small $N_{\rm vec}$, and 
is saturated at around $N_{\rm vec}\sim 30$-$100$ 
to reach $\sim$70-80 \% of the theoretical peak performance.  
The values of the two systems are very close at a large $N_{\rm vec}$ 
in contrast to rather different behavior 
at a smaller $N_{\rm vec}$. 

Although the matrix-vector and matrix-matrix methods are 
identical in mathematics, they are quite different 
in performance. 
Memory access, the major bottleneck of modern 
computer systems, differentiates the methods from each other. 
Now we consider 
a matrix of size $n\times n$ and a vector of size $n$
and estimate the number of arithmetic operations and memory accesses 
involving them. 
Since a matrix-times-vector operation 
needs $2n^2$ floating-point 
operations and $n^2+n$ memory accesses, 
the computational intensity defined by their ratio 
is $\sim 2$. On the other hand, 
the computational intensity 
for a matrix-times-matrix operation becomes $n$, much larger 
than that of the matrix-vector operation for a sufficiently large $n$. 
More specifically, 
the matrix-times-matrix operation can be 
designed so that most of the CPU time can be involved in 
arithmetic operations rather than memory access 
as is implemented in numerical libraries such as MKL
and SSL II.   
See, for instance, \cite{hpc} for more detailed 
analyses of the performance of basic linear algebra operations 
in terms of computer architecture. 

We also consider the performance of parallel processes. 
We take an example where the 31,250 $^t\tilde{\rho}\tilde{v}\tilde{\rho}$ 
operations are divided into 32 MPI processes running on a 
4 node $\times$ 2 CPU $\times$ 4 core Xeon X5570 system. 
The matrix-matrix method with $N_{\rm vec}=100$ reaches 
8.5 GFLOPS/core, which is rather close to the 
9.1 GFLOPS achieved by the single process. In contrast, 
for the matrix-vector method, 
the parallel performance is reduced to 1.5 GFLOPS/core 
from the single-process performance of 3.1 GFLOPS. 
This difference is also accounted for by the memory access: 
since the memory bandwidth is shared by all the CPU cores 
on the board, the effective bandwidth defined by the bandwidth 
per process or thread is reduced 
for parallel processes. This reduction of the effective bandwidth 
leads to  
the reduction of the performance 
particularly for the processes involving heavy memory access like 
the matrix-vector operation. 
Thus, the matrix-matrix method is superior 
to the matrix-vector method 
not only in absolute performance but also 
in parallel efficiency  
because of less memory demanding formalism. 

\subsection{Towards larger calculations}

\begin{figure}[t]
\begin{center}
\includegraphics[width=8.0cm,clip]{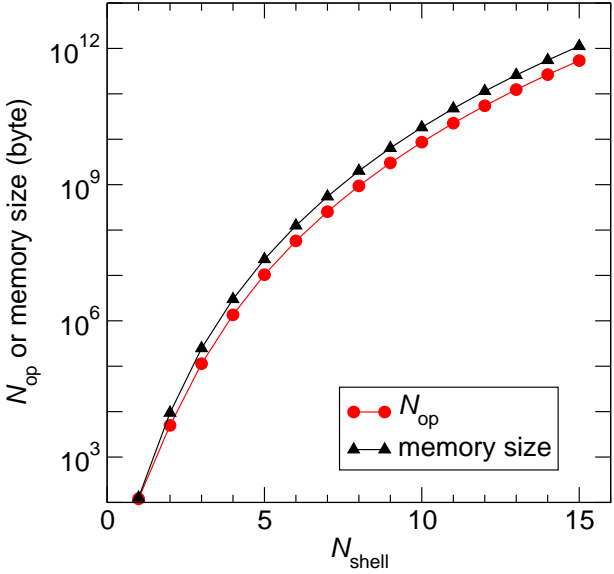}
\caption{ 
The number of elementary floating-point operations $N_{\rm op}$ 
and memory size in byte 
needed in the matrix-vector method as 
a function of $N_{\rm shell}$. 
Double precision data are assumed for estimating the memory size. 
}
\label{fig:nshell}
\end{center}
\end{figure}

The benchmark calculation is carried out 
for the model space $N_{\rm shell}=5$, where $N_{\rm shell}$ 
stands for 
the number of harmonic oscillator major shells included. 
Although this model space 
is large for the shell-model calculation, 
it is not large enough for the density-functional calculation.  
Hence, in view of possible application to modern 
density-functional calculations, 
we examine how computational requirements  
change as the model space is enlarged. 

Figure \ref{fig:nshell} shows the increase of $N_{\rm op}$ 
defined as the number of elementary floating-point operations 
for computing a single $\langle V \rangle$ of Eq.~(\ref{eq:vcal})
with the matrix-vector method. 
In estimating the number, it is assumed that a multiplication 
and an addition of two complex numbers need six and two 
floating-point operations, respectively, while a multiplication 
between a real number and a complex number needs 
two floating-point operations. 
Figure \ref{fig:nshell} indicates that 
$N_{\rm op}$ increases roughly exponentially with $N_{\rm shell}$ 
but that the slope decreases. 
As a result, the computational time for $N_{\rm shell}=10$ 
is, for instance, $\sim 10^3$ times larger than the one for 
$N_{\rm shell}=5$ 
if the effective performance shown in Fig.~\ref{fig:performance} 
is unchanged. 
This assumption is reasonable because 
the effective performance for a multiplication of large matrices 
is known to be kept high. 
Indeed, we have confirmed that almost the same 
performance is obtained for $N_{\rm shell}=6$. 

On the practical side, memory size could be a problem. 
As demonstrated in Fig.~\ref{fig:nshell}, gigabytes of memory 
are required to store the two-body matrix elements 
$\bar{v}_{l_1l_2,l_3l_4}$ for $N_{\rm shell}\ge 8$. 
However, this restriction due to the memory size can be relaxed 
when matrices $\tilde{v}(\Delta m, \Delta m)$ having different 
$\Delta m$ are distributed over different nodes. 
In the case of $N_{\rm shell}=10$, since the largest matrix size is 
12,444, the maximum memory size is reduced 
to $\sim 1$ GB.  
Finally, it should be noted that while the memory size 
needed for the calculation of 
multiple $\langle V \rangle$'s using the matrix-matrix method 
is almost unchanged from that of single $\langle V \rangle$, 
the number of floating-point operations is multiplied by 
$N_{\rm vec}$, i.e., the number of vectors bound (see Eq.~(\ref{eq:vtheta})). 
Thus, the ratio of the number of memory access to the number of 
operations decreases accordingly, as discussed in
Sec.~\ref{sec:result}. 

\section{Summary}
\label{sec:summary}

We have presented an efficient numerical method for computing 
Hamiltonian matrix elements between non-orthogonal 
Slater determinants, motivated by recent findings that 
a superposition of non-orthogonal Slater determinants is 
a very effective way to solve a many-body problem. 
The most computationally demanding is the computation of 
a four-fold loop 
$\langle V \rangle 
=\sum_{l_1l_2l_3l_4}\rho_{l_3l_1}\,\bar{v}_{l_1l_2,l_3l_4}\,\rho_{l_4l_2}$, 
where $\bar{v}_{l_1l_2,l_3l_4}$ is a sparse array due to 
the symmetries of the Hamiltonian. 
While indirectly indexed arrays are often introduced 
for treating a sparse matrix, the performance of the method has been 
measured to be much lower than the theoretical peak performance. 
In order to fit a formula of calculating $\langle V \rangle$ to fast
computation, 
its key part is 
transformed into a multiplication of a dense matrix and a vector 
for a single $\langle V \rangle$ calculation. 
This formula is also transformed into a multiplication of dense matrices 
for multiple $\langle V \rangle$ calculations.  
The method based on the matrix-matrix multiplication 
attains as much as $\sim$80\% of the theoretical peak performance 
on actual systems. Its high performance is accounted for 
by its high computational intensity, i.e., a large 
ratio of floating-point operations to memory accesses. 
Since from the hardware side it is predicted that the Byte/FLOP rate of future systems 
will be decreased \cite{sarkar}  
because of rapid increase of the number of CPU cores compared to memory bandwidth, 
numerical methods should be developed so that the computational
intensity 
can be higher as achieved by the present method. 

\section*{Acknowledgement}
One of the authors (Y.U.) thanks Prof. J. Dobaczewski 
for fruitful discussions during his stay at the European Centre 
for Theoretical Studies in Nuclear Physics and Related Areas (ECT*). 
This work was in part supported by MEXT Grant-in-Aid for 
Young Scientists (B) (21740204), 
for Scientific Research on Innovative Areas
(No. 20105003), 
for Scientific Research (A) (20244022, 23244049), 
and the HPCI Strategic Program of MEXT. 
This work is a part of 
the RIKEN-CNS joint research project on large-scale 
nuclear-structure calculations. 
The  numerical calculation was carried 
out on the BX900 and FX1 supercomputers at the Japan Atomic 
Energy Agency. 



\appendix
\section{Derivation of the formulae for calculating 
the norm and Hamiltonian overlaps}
\label{app:kernel}

In this appendix, we derive the formulae for calculating the norm and Hamiltonian 
overlaps given by 
Eqs. (\ref{eq:norm_kernel}) and (\ref{eq:hamiltonian_kernel}). 
In the following, it is convenient to introduce unoccupied states 
\begin{equation}
\label{eq:unoccupied}
a^{\dag}_m(q) = \sum_{l}^{N_s}\tilde{D}(q)_{lm}c^{\dag}_l . 
\end{equation}
Hereafter, the occupied and unoccupied states are labeled 
by the indices $i,j$ and $m,n$, respectively. Using
Eqs.~(\ref{eq:occupied}) and (\ref{eq:unoccupied}), 
the creation operator of a single-particle basis state $c^{\dag}_l$ is 
written by $a^{\dag}_i(q)$ and $a^{\dag}_m(q)$ as 
\begin{equation}
\label{eq:c_from_a}
c^{\dag}_l = \sum_i D(q)_{li}^* a^{\dag}_i + \sum_m \tilde{D}(q)_{lm}^*
a^{\dag}_m .
\end{equation}
The anticommutation relation $\{ c_l, c_{l'}^{\dag}\} = \delta_{ll'}$ 
leads to 
\begin{equation}
\label{eq:d_orth}
D(q)D(q)^{\dag} + \tilde{D}(q)\tilde{D}(q)^{\dag} = I, 
\end{equation}
where $I$ is the identity matrix. 
The creation operator $a^{\dag}_j(q')$ can be expressed as a linear 
combination of $a^{\dag}_i(q)$ and $a^{\dag}_m(q)$ 
by using Eqs.~(\ref{eq:occupied}) and (\ref{eq:c_from_a}): 
\begin{equation}
a^{\dag}_j(q') = \sum_i E_{ij} a^{\dag}_i(q) + \sum_m \tilde{E}_{mj}
 a^{\dag}_m(q), 
\end{equation}
where $E$ and $\tilde{E}$ are given by 
$E=D(q)^{\dag}D(q')$ and $\tilde{E}=\tilde{D}(q)^{\dag}D(q')$,
respectively. 

\subsection{Norm overlap}
The overlap between $\left| \Phi(q') \rangle\right.$ and 
$\left| \Phi(q) \rangle\right.$ is calculated as 
\begin{equation}
\begin{array}{l}
   \langle \Phi(q') | \Phi(q) \rangle \\
 = 
\displaystyle{
\langle - | \prod_{i'=N_p}^{1} a_{i'}(q')\prod_{i=1}^{N_p} a^{\dag}_i(q) | -\rangle
} \\
  =  
\displaystyle{
\langle - | \prod_{i'=N_p}^{1}  \sum_j E_{ji'}^{*} a_{j}(q) 
\prod_{i=1}^{N_p} a^{\dag}_i(q) | -\rangle
} \\
  = 
\displaystyle{
\sum_{\sigma \in S_{N_p}} E_{\sigma(1)1}^{*}\cdots 
E_{\sigma(N_p)N_p}^{*}} \\ \times
 \langle - |  a_{\sigma(N_p)}\cdots 
a_{\sigma(1)} a^{\dag}_1  \cdots a^{\dag}_{N_p} | - \rangle
 \\
  =  \det{E^{\dag}} \\
  =  \det \left( {D(q')^{\dag}D(q)} \right) , 
\end{array}
\end{equation}
where $S_{N_p}$ stands for the symmetric group of degree $N_p$. 

\subsection{Hamiltonian overlap}

According to Thouless' theorem \cite{thouless}, 
any Slater determinant $\left| \Phi(q') \rangle\right.$
that is not orthogonal to a Slater determinant $\left| \Phi(q)
\rangle\right.$ can be expressed as 
\begin{equation}
\label{eq:thouless}
| \Phi(q') \rangle = N e^{\hat{Z}} | \Phi(q) \rangle, 
\end{equation}
where $\hat{Z} = \sum_{i,m} Z_{mi}a^{\dag}_m a_i$. 
The normalization constant $N$ is given by 
$N=\langle \Phi(q) | \Phi(q') \rangle$. 
Using Eq.~(\ref{eq:thouless}), a general matrix element between 
$\left| \Phi(q') \rangle\right.$ and $\left| \Phi(q) \rangle\right.$
is 
\begin{equation}
\label{eq:d_expression}
\begin{array}{cl}
& \langle  \Phi(q')  | c^{\dag}_{l_1}\cdots c^{\dag}_{l_p} 
c_{k_1} \cdots c_{k_q} 
 | \Phi(q)\rangle \\
= & \langle \Phi(q') | \Phi(q) \rangle \langle \Phi(q) | \bar{d}_{l_1} \cdots
 \bar{d}_{l_p} d_{k_1} \cdots d_{k_q} | \Phi(q) \rangle , 
\end{array}
\end{equation}
where $\bar{d_l}$ and $d_l$ are defined by 
$\bar{d_l} = e^{\hat{Z}^{\dag}} c^{\dag}_l e^{-\hat{Z}^{\dag}} $ and 
$d_l = e^{\hat{Z}^{\dag}}  c_l e^{-\hat{Z}^{\dag}} $, respectively. 
When the creation operator $b^{\dag}$ is defined by  
$b^{\dag}_m = a^{\dag}_m$ and 
$b^{\dag}_i = a_i$, $| \Phi(q) \rangle$ is regarded as 
vacuum: 
\begin{equation}
\label{eq:b_vacuum}
b_l | \Phi(q) \rangle = 0
\end{equation}
for any single-particle state $l$.  Hence, it is useful to represent 
$\bar{d_l}$ and $d_l$ with $b^{\dag}$ and $b$. 
Hereafter, $D(q)$ issimply written as $D$ when 
no confusion is possible. 
Using the Baker-Hausdorff formula, it is straightforward to 
derive
\begin{equation}
\begin{array}{rcl}
 \bar{d_l} & = & 
\displaystyle{
\sum_i \left( D_{li}^{*} + \sum_{m} Z_{mi}^{*}\tilde{D}_{lm}^{*} \right)
b_i + \sum_m \tilde{D}_{lm}^{*} b^{\dag}_m 
}
\\
d_l & = & 
\displaystyle{
\sum_i D_{li} b^{\dag}_i + \sum_m \left( 
\tilde{D}_{lm} - \sum_i Z_{mi}^{*} D_{li}
\right) b_m 
} .
\end{array}
\end{equation}

Wick's theorem \cite{wick} is helpful to calculate the right hand side of 
Eq.~(\ref{eq:d_expression}). To use this theorem, the contraction of operators 
$U$ and $V$ defined as 
$ U^{\bullet}V^{\bullet}= UV- \colon UV\colon$
is needed, where $\colon UV \colon$ stands for the normal ordered product 
of $UV$ concerning $b^{\dag}$ and $b$. 
Different expressions $ U^{\bullet\bullet}V^{\bullet\bullet}$, 
$ U^{\bullet\bullet\bullet}V^{\bullet\bullet\bullet}$  {\it etc.} are 
also used for  the contraction $ U^{\bullet}V^{\bullet}$ in order to 
specify the pair of operators considered. 
This definition leads to the following contractions 
\begin{equation}
\begin{array}{rcl}
\bar{d}_p^{\bullet} d_q^{\bullet} & = & 
\left( D(D^{\dag}+Z^{\dag}\tilde{D}^{\dag})\right)_{qp}
\\
d_p^{\bullet} \bar{d}_q^{\bullet} & = & 
\left( (\tilde{D} - DZ^{\dag})\tilde{D}^{\dag}\right)_{pq} \\
\bar{d}_p^{\bullet} \bar{d}_q^{\bullet} & = & 0 \\
d_p^{\bullet} d_q^{\bullet} & = & 0 .
\end{array}
\end{equation}
The density matrix defined in Eq.~(\ref{eq:density}) 
is identical with the contraction (see Eq.~(\ref{eq:d_expression})):  
\begin{equation}
\label{eq:rho_z}
\rho_{ll'} = \bar{d}_{l'}^{\bullet} d_l ^{\bullet}
= \left( D(D^{\dag}+Z^{\dag}\tilde{D}^{\dag})\right)_{ll'} . 
\end{equation}
The explicit form of the matrix $Z$ can be derived from the 
condition
\begin{equation}
a^{\dag}_i(q') | \Phi(q') \rangle = Ne^{\hat{Z}}\left( e^{-\hat{Z}} a^{\dag}_i(q') 
e^{\hat{Z}} \right)
| \Phi(q) \rangle  = 0 .
\end{equation}
After some lengthy calculations, it is proved that this is satisfied 
when $Z$ is taken to be 
\begin{equation}
\label{eq:z}
Z = \tilde{E}E^{-1} = \tilde{D}(q)^{\dag} D(q')
 (D(q)^{\dag}D(q'))^{-1}. 
\end{equation}
Thus, the expression 
\begin{equation}
\rho = D(q)\left( D(q')^{\dag}D(q)\right)^{-1}D(q')^{\dag}
\end{equation}
is obtained by substituting Eq.~(\ref{eq:z}) for Eq.~(\ref{eq:rho_z}) 
and using Eq.~(\ref{eq:d_orth}). 

In the case of a two-body operator $\bar{d}_{l_1}  \bar{d}_{l_2} d_{l_4}
d_{l_3}$, 
Wick's theorem leads to 
\begin{equation}
\label{eq:contr_tb}
\begin{array}{cl}
 & \bar{d}_{l_1}  \bar{d}_{l_2} d_{l_4} d_{l_3} \\ 
= & \colon \bar{d}_{l_1}\bar{d}_{l_2} d_{l_4} d_{l_3}\colon  \\
   + & \colon \bar{d}_{l_1}^{\bullet}\bar{d}_{l_2}^{\bullet} d_{l_4}
 d_{l_3}\colon +  \colon \bar{d}_{l_1}^{\bullet}\bar{d}_{l_2}d_{l_4}^{\bullet}
 d_{l_3}\colon + \cdots  \\
 + & \colon \bar{d}_{l_1}^{\bullet}\bar{d}_{l_2}^{\bullet} d_{l_4}^{\bullet\bullet}
 d_{l_3}^{\bullet\bullet}\colon  +
\colon \bar{d}_{l_1}^{\bullet}\bar{d}_{l_2}^{\bullet\bullet} d_{l_4}^{\bullet}
 d_{l_3}^{\bullet\bullet}\colon  + 
\colon \bar{d}_{l_1}^{\bullet}\bar{d}_{l_2}^{\bullet\bullet}d_{l_4}^{\bullet\bullet}
 d_{l_3}^{\bullet}\colon  . 
\end{array}
\end{equation}
While the last three terms of the right hand side 
of Eq.~(\ref{eq:contr_tb}) are $c$-numbers,  the other terms 
include $b$ and/or $b^{\dag}$ in the normal order and produce vanishing 
diagonal matrix elements for $| \Phi(q) \rangle$
because of Eq.~(\ref{eq:b_vacuum}).
Thus, 
the general matrix element of a two-body operator
$c^{\dag}_{l_1}c^{\dag}_{l_2}c_{l_4}c_{l_3}$
between $| \Phi(q) \rangle$ and $| \Phi(q') \rangle$
is given by Eq.~(\ref{eq:d_expression}): 
\begin{equation}
\label{eq:gen_tb}
\begin{array}{cl}
& \langle  \Phi(q')  | c^{\dag}_{l_1}c^{\dag}_{l_2} 
c_{l_4}  c_{l_3} 
 | \Phi(q)\rangle \\
= & \langle \Phi(q') | \Phi(q) \rangle \langle \Phi(q) | \bar{d}_{l_1} 
 \bar{d}_{l_2} d_{l_4} d_{l_3} | \Phi(q) \rangle \\
= & \langle \Phi(q') | \Phi(q) \rangle ( 
\colon \bar{d}_{l_1}^{\bullet}\bar{d}_{l_2}^{\bullet} d_{l_4}^{\bullet\bullet}
 d_{l_3}^{\bullet\bullet}\colon  +
\colon \bar{d}_{l_1}^{\bullet}\bar{d}_{l_2}^{\bullet\bullet} d_{l_4}^{\bullet}
 d_{l_3}^{\bullet\bullet}\colon  \\ &  +
\colon \bar{d}_{l_1}^{\bullet}\bar{d}_{l_2}^{\bullet\bullet}d_{l_4}^{\bullet\bullet}
 d_{l_3}^{\bullet}\colon  
) \\ 
= & \langle \Phi(q') | \Phi(q) \rangle
( \rho_{l_3 l_1}\rho_{l_4 l_2} - \rho_{l_4 l_1}\rho_{l_3 l_2}), 
\end{array}
\end{equation}
where 
$\colon \bar{d}_{l_1}^{\bullet}\bar{d}_{l_2}^{\bullet} d_{l_4}^{\bullet\bullet}
 d_{l_3}^{\bullet\bullet}\colon =0$, 
$\colon \bar{d}_{l_1}^{\bullet}\bar{d}_{l_2}^{\bullet\bullet} d_{l_4}^{\bullet}
 d_{l_3}^{\bullet\bullet}\colon =
-\colon \bar{d}_{l_1}^{\bullet}d_{l_4}^{\bullet}\bar{d}_{l_2}^{\bullet\bullet} 
d_{l_3}^{\bullet\bullet}\colon 
= -\rho_{l_4 l_1}\rho_{l_3 l_2}$, and 
$\colon \bar{d}_{l_1}^{\bullet}\bar{d}_{l_2}^{\bullet\bullet}d_{l_4}^{\bullet\bullet}
 d_{l_3}^{\bullet}\colon = 
\colon \bar{d}_{l_1}^{\bullet}d_{l_3}^{\bullet}\bar{d}_{l_2}^{\bullet\bullet} 
d_{l_4}^{\bullet\bullet}\colon 
= \rho_{l_3 l_1}\rho_{l_4 l_2}$
are used. It is noted that 
any transposition of two operators in the contraction changes the sign 
(see Rule C'' in \cite{wick}). 
This gives Eq.~(\ref{eq:hamiltonian_kernel})  straightforwardly 
for antisymmetrized two-body matrix elements satisfying 
$\bar{v}_{l_1l_2,l_3l_4} =- \bar{v}_{l_1l_2,l_4l_3}$.




\bibliographystyle{elsarticle-num}
\bibliography{<your-bib-database>}

\begin{thebibliography}{00}



\bibitem{ring} For instance, 
P.~Ring and P.~Schuck, The Nuclear Many-Body Problem, 
Springer, 1980. 

\bibitem{gcm}
D.L.~Hill and J.A.~Wheeler, Phys. Rev. {\bf 89} (1953) 1102; 
J.J.~Griffin and J.A.~Wheeler, Phys. Rev. {\bf 108} (1957) 311. 

\bibitem{bender03}
M.~Bender, P.-H.~Heenen, and P.-G.~Reinhard, 
Rev. Mod. Phys. {\bf 75} (2003) 121. 

\bibitem{bender06}
M.~Bender, G.F.~Bertsch, and P.-H.~Heenen, 
Phys. Rev. C {\bf 73} (2006) 034322. 

\bibitem{sabbey07}
B.~Sabbey, M.~Bender, G.F.~Bertsch, and P.-H.~Heenen, 
Phys. Rev. C {\bf 75}  (2007) 044305. 

\bibitem{robledo11}
L.M.~Robledo and G.F.~Bertsch, 
Phys. Rev. C {\bf 84}  (2011) 054302. 

\bibitem{mcsm}
T.~Otsuka, M.~Honma, T.~Mizusaki, N.~Shimizu, and 
Y.~Utsuno, Prog. Part. Nucl. Phys. {\bf 47} (2001) 319. 

\bibitem{vampir}
K.W.~Schmid, Prog. Part. Nucl. Phys. {\bf 52} (2004) 565. 

\bibitem{puddu}
G.~Puddu, J. Phys. G {\bf 32} (2006) 321. 

\bibitem{extrapolation}
N.~Shimizu, Y.~Utsuno, T.~Mizusaki, T.~Otsuka, T.~Abe, and M.~Honma,
Phys. Rev. C {\bf 82} (2010) 061305(R). 

\bibitem{koch}
H.~Koch and E.~Dalgaard, Chem. Phys. Lett. {\bf 212} (1993) 193. 

\bibitem{tomita}
N.~Tomita, S.~Ten-no, and Y.~Tanimura, 
Chem. Phys. Lett. {\bf 263} (1996) 687. 

\bibitem{morgon}
N.H.~Morgon, J. Phys. Chem. A {\bf 102} (1998) 2050. 

\bibitem{scuseria}
G.E.~Scuseria, C.A.~Jim\'{e}nez-Hoyos, T.M.~Henderson, K.~Samanta, 
and J.K.~Ellis, J. Chem. Phys. {\bf 135} (2011) 124108. 

\bibitem{bm} For instance, A.~Bohr and B.R.~Mottelson, Nuclear
	Structure, Vol. 1, Benjamin, 1969. 

\bibitem{blas} 
C.L.~Lawson, R.J.~Hanson, D.~Kincaid, and F.T.~Krogh, 
ACM Trans. Math. Soft. {\bf 5} (1979) 308; 
J.J.~Dongarra, J.~Du~Croz, S.~Hammarling, and R.J.~Hanson, 
ACM Trans. Math. Soft. {\bf 14} (1988) 1; 
J.J.~Dongarra, J.~Du~Croz, S.~Hammarling, and R.J.~Hanson, 
ACM Trans. Math. Soft. {\bf 14} (1988) 18; 
J.J.~Dongarra, J.~Du~Croz, I.S.~Duff, and S.~Hammarling, 
ACM Trans. Math. Soft. {\bf 16} (1990) 1; 
J.J.~Dongarra, J.~Du~Croz, I.S.~Duff, and S.~Hammarling, 
ACM Trans. Math. Soft. {\bf 16} (1990) 18.

\bibitem{hpc}
K.R.~Wadleigh and I.L.~Crawford, 
Software Optimization for High Performance Computing: 
Creating Faster Applications, Prentice Hall, 2000. 

\bibitem{sarkar} 
V.~Sarkar, W.~Harrod, and A.E.~Snavely, 
J. Phys.: Conf. Ser. {\bf 180} (2009) 012045. 

\bibitem{thouless}
D.J.~Thouless, Nucl. Phys. {\bf 21} (1960) 225. 

\bibitem{wick}
G.C.~Wick, Phys. Rev. {\bf 80} (1950) 268. 

\end{thebibliography}



\end{document}